\documentclass[12pt,preprint]{aastex}

\def\etal{{et~al}\hbox{.}}

\slugcomment{Submitted to ApJ: March 21, 2001}

\shorttitle{Mass Function of Low Mass Galaxies in CDM}
\shortauthors{Chiu, Gnedin, \& Ostriker}

\begin{document}


\title{
The Expected Mass Function for \\
Low Mass Galaxies in a CDM Cosmology:\\
Is There a Problem?}


\author{Weihsueh A. Chiu\altaffilmark{1}}
\affil{Joseph Henry Laboratories, Department of Physics, \\
        Princeton University, Princeton, NJ 08544}
\email{chiu@astro.princeton.edu}

\author{Nickolay Y. Gnedin}
\affil{Center for Astrophysics and Space Astronomy, \\
University of Colorado, Boulder, CO 80309}
\email{gnedin@casa.colorado.edu}

\and

\author{Jeremiah P. Ostriker}
\affil{Princeton University Observatory, Princeton, NJ 08544}
\email{jpo@astro.princeton.edu}


\altaffiltext{1}{present address:  
U.S. Environmental Protection Agency, 
Mail Code 6608J, 1200 Pennsylvania Ave. NW,
Washington, DC 20460}


\begin{abstract}
It is well known that the mass function for \emph{halos} in 
CDM cosmology is a 
relatively steep power law ($n(M) \propto M^{-2}$) for low
masses, possibly too steep to be consistent with observations.
But how steep is the \emph{galaxy} mass function?
We have analyzed the stellar and gas mass functions of 
the first massive luminous objects 
formed in a $\Lambda$CDM universe, as calculated
in the numerical simulation described in Gnedin (2000ab).
Our analysis indicates that, as suspected, the relationship between
luminous or gaseous matter and total (primarily dark matter) 
mass is not a linear
scaling relation --- i.e., light does not trace mass.  
While the dark matter mass function
is steep, the stellar and gas mass functions are flatter for
low mass objects.
The stellar mass function is consistently 
flat at the low mass end.  
Moreover, while the gas mass function follows the dark matter mass function
until reionization at $z\sim 7$, between $z=7$ and $z=4$, 
the gas mass function also flattens considerably at the low mass end.
At $z=4$, the gas and stellar mass functions
are fit by a Schechter function with 
$\alpha \approx -1.2\pm 0.1$, significantly shallower than
the dark matter halo mass function and consistent with some recent
observations.  The baryonic mass functions are shallower 
because (a) the dark matter halo mass (differential) function 
is consistent with the Press-Schechter
formulation at low masses $n(M) \propto M^{-2}$ and
(b) heating/cooling and ionization processes appear to cause 
baryons to collect in halos with the relationship $M_b \propto M_d^4$
at low masses.  Combining (a) and (b) gives $n(M_b) \propto M_b^{-5/4}$,
comparable to the simulation results. 
Thus, the well known observational fact that low mass galaxies are
underabundant as compared to expectations from numerical 
dark matter simulations
or Press-Schechter modeling of CDM universes 
emerges naturally from these results,
implying that perhaps no ``new physics'' beyond the 
standard model is needed.
\end{abstract}


\keywords{cosmology --- galaxies: abundances, dwarf, formation}


\section{Introduction}

Recent interest in modifying the standard $\Lambda$CDM theory of the
origin of structure in the universe has been motivated
largely by observations on sub-galactic scales.  
In particular, the most straight-forward interpretation 
of the results of theoretical calculations appear to
be in conflict with the observed characteristics
of spheroidal galaxies (i.e., bulges of spirals and 
small to moderate ellipticals) with respect to their individual
physical properties, luminosity function, 
and clustering properties (Sellwood and Kosowsky 2000,
Klypin \etal\ 2000, Navarro and Steinmetz 2000).
These facts have led some authors
to turn to alternative models such as quantum isocurvature (Peebles 1999ab),
Self-Interacting Dark Matter \citep{SS00},
and Warm Dark Matter \citep{Bode01}. 

However, while measurements such as those of the cosmic microwave background
radiation can be directly computed from theory in a relatively
straight-forward manner, the abundance and clustering of luminous
galaxies cannot at the present time be directly calculated 
with any certainty.  Such calculations are hampered by the classic
problem of relating light to mass.  
The simplest way to relate dark matter density to galaxy counts is to assume 
a proportionality in the fluctuations.  This has led to the invention of
the ``spatial bias parameter $b$.''  
With the increasing sophistication of both
theory and observations, we have discovered that bias must depend on 
both length scale and galactic type (because different types of galaxies
have different clustering properties) and possibly on both local density
and temperature \citep{Blanton00a}. 

But what is the relation between dark matter halos and galaxies 
as a function of mass?  Is
it plausible to assume that each dark matter halo contains a galaxy having
stellar mass and light proportional to the dark matter mass?  The earliest
phenomenological treatments explored this possibility, but observationally
it is clearly false.  For example, 
the most massive ($\sim 10^{15} M_\odot$) identifiable
dark matter halos --- found from large angle arcs and multiply imaged 
galaxies --- correspond to clusters of distinct galaxies and do not contain
supergiant galaxies of corresponding baryonic mass 
($10^{13} - 10^{14} M_\odot$).

For dark matter halos less than or comparable to $\sim 10^{12} M_\odot$, the 
prevailing belief is that they will typically contain one ``galaxy'' and 
several satellites, but whether this description is accurate (and 
informative rather than definitional) is not known very securely.  In the
semianalytic treatment of galaxy formation (Somerville and Primack 1999, 
Lobo and Guiderdoni 1999, Chiu and Ostriker 2000) halos are identified
with galaxies by a set of prescriptions that, overall, appears to correspond
to reality.  But, hydrodynamic simulations tend in this mass range to find
several galaxies orbiting within a common halo \citep{Cen00}. 

For still lower mass galaxies (and presumably lower mass halos), we know
that the mass/light ratio becomes large, implying that galaxy formation was 
inefficient for small systems.  Thus, we are led to ask, could the 
physical bias against low mass systems naturally produce a mass function
for galaxies significantly flatter than that of the dark matter halo mass
function?  If so, some of the apparent discrepancies between CDM
cosmologies and observations might disappear. 

It may be that some of the other discrepancies, such as those
that depend on lensing \citep{LO00}, 
shape (Moore \etal\ 1999b, Miralda-Escud\'e 2000, Navarro and Steinmetz 2000), 
or small scale potential fluctuations 
\citep{Moore99a}, will remain
and require serious reexamination of the CDM paradigm.  But the
driving reason for reexamination --- the much smaller than expected number
density of small galaxies --- might simply be a result of well understood
physical properties.  This has been suggested already (Silk 1977, Lobo and
Guiderdoni 1999) on the basis of semianalytical arguments.  
The purpose of this paper
is to see if the mass function of low mass galaxies (``faint'' galaxies) 
is significantly flatter
than that for low mass halos, and the evidence from the hydrodynamic 
simulation examined (Gnedin 2000ab) is \emph{yes}.

The organization of this paper is as follows.
In \S~\ref{sec:simulation} we describe the hydrodynamical simulation used.
In \S~\ref{sec:sph_massfunc} we 
show our analysis of the mass functions of the bound 
systems in the simulation.
We discuss the why the galaxy mass function is relatively flat 
in \S~\ref{sec:discussion},
and conclude in \S~\ref{sec:conclusions}.

\section{The Hydrodynamical Simulation}
\label{sec:simulation}

The simulation code used here is based 
on Smoothed Lagrangian Hydrodynamics and
is fully described in \citet{G00a}.  It is 
an extension of that reported on in 
Gnedin (1995, 1996), \citet{GB96}, and \citet{GO97}.
In brief, this simulation contains $128^3$ dark matter particles, a baryonic
mesh of the same size, has a spatial resolution of 1.0 kpc/$h$, 
and a box size of $4^3$ (Mpc/$h$)$^{3}$.  The baryonic mass resolution
is $10^{5.7} M_{\odot}$. 
In addition, the code includes heuristic star formation, with
the creation of stellar ``particles'' whose dynamics only depends
on gravity.  Finally, reionization by these ``stars'' is modeled 
with the local optical depth approach which is able to approximately
follow three-dimensional radiative transfer.  Because the present day
nonlinear scale is larger than the box size, the simulation can not
be continued to the present, and was stopped at a redshift $z=4$.
Bound objects at each epoch are identified through use of the 
DENMAX \citep{bertgelb91} algorithm.  The simulation run analyzed 
here is the same as the ``production run'' analyzed in Gnedin (2000ab). 
The cosmological parameters are $\Omega_0 = 0.3$, $\Omega_b = 0.04$,
$\Omega_\Lambda=0.7$, $h=0.7$, $n=1$, and $\sigma_8=0.9$ (COBE normalized).

Using results of a similar code, 
\citet{GO97} showed that ionization is largely complete for 
a standard $\Lambda$CDM model at $z=7$.  In addition, 
\citet{GNO00} recently showed that when 
the effects of gas cooling and star formation are included, 
massive bulges are able to collapse to densities exceeding 
that of the dark matter, 
resulting in structures that resemble observed objects in 
their sizes, shapes, and density profiles.

\section{Mass Functions of Spheroidal Galaxies}
\label{sec:sph_massfunc}

Here, we examine the mass functions of the dark and baryonic 
components of the bound objects formed in our hydrodynamical simulation.  
From the simulation results, 
we calculate the mass function $n_i(>M_i),$ the number of 
objects with mass in component $i > M_i$, where $i=$dark matter (d), 
gas (g), or stellar particles (s).  
If dark matter, gas, and stellar 
particles all trace each other, then 
\begin{equation}
n_i(>M_i) = 
n_j(>M_j)
\label{eq:numscale}
\end{equation}
with the mass scaling 
\begin{equation}
	\frac{M_i}{M_j} = \frac{f_i}{f_j},
\label{eq:linmassscale}
\end{equation}
where $f_i$ is the fraction of the mass of the universe
in species $i$.  The dark matter fraction $f_d=\Omega_d/\Omega_0=0.867$ 
is universal.  Baryons are either in the form of
gas or stellar particles, so that $f_g+f_s=0.133$.  
The fraction of mass in stellar particles is given in 
Table~\ref{tab:meanmassfrac} as a function of redshift.  At all
times, less than 4\% 
of the baryons are in the form of stellar particles, so $f_g$ remains
approximately constant.

\begin{figure}
\plotone{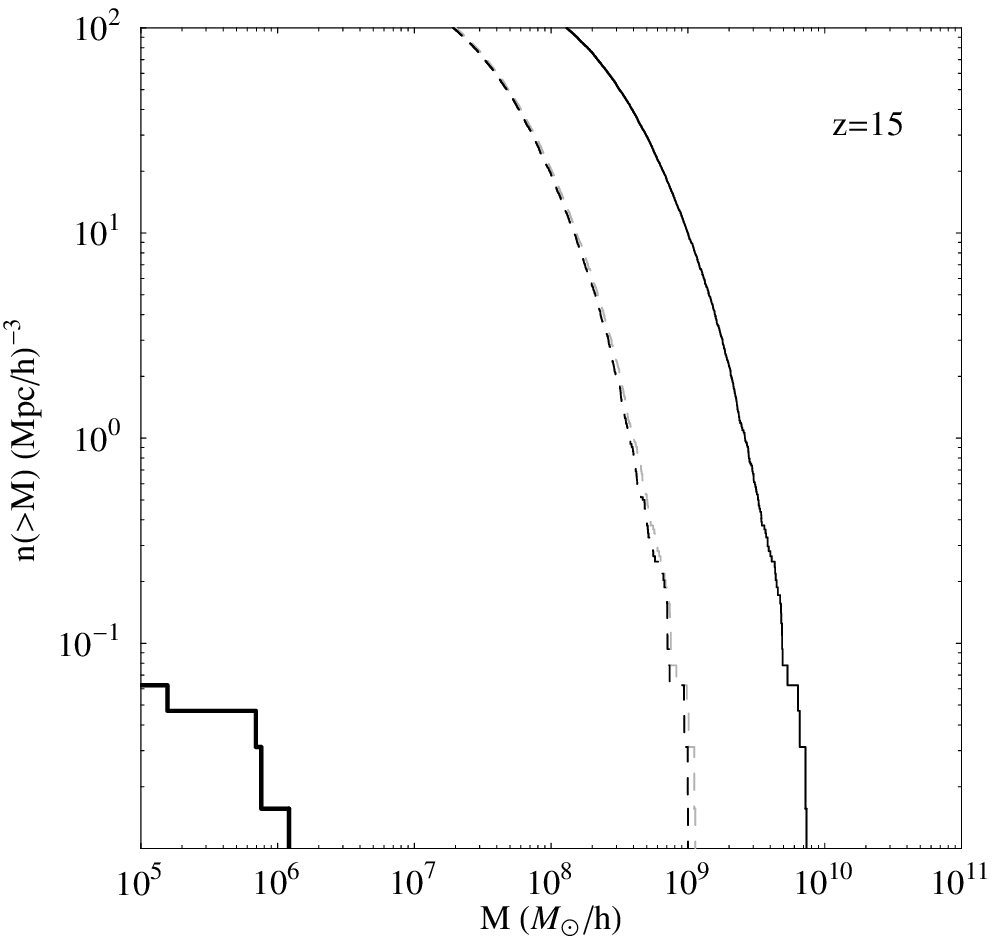}
\caption{Mass functions for bound objects 
at $z=15$, before reionization.  
Shown are the dark matter (thin solid), gas (thin dashed), and 
stellar particles (thick solid).  Also shown (dashed grey) 
is the dark matter mass function with the mass scaled by the universal 
ratio of baryons to dark matter $\Omega_{\rm b}/\Omega_{\rm d}$.
}
\label{fig:massfunc_one}
\end{figure}

\begin{figure}
\plotone{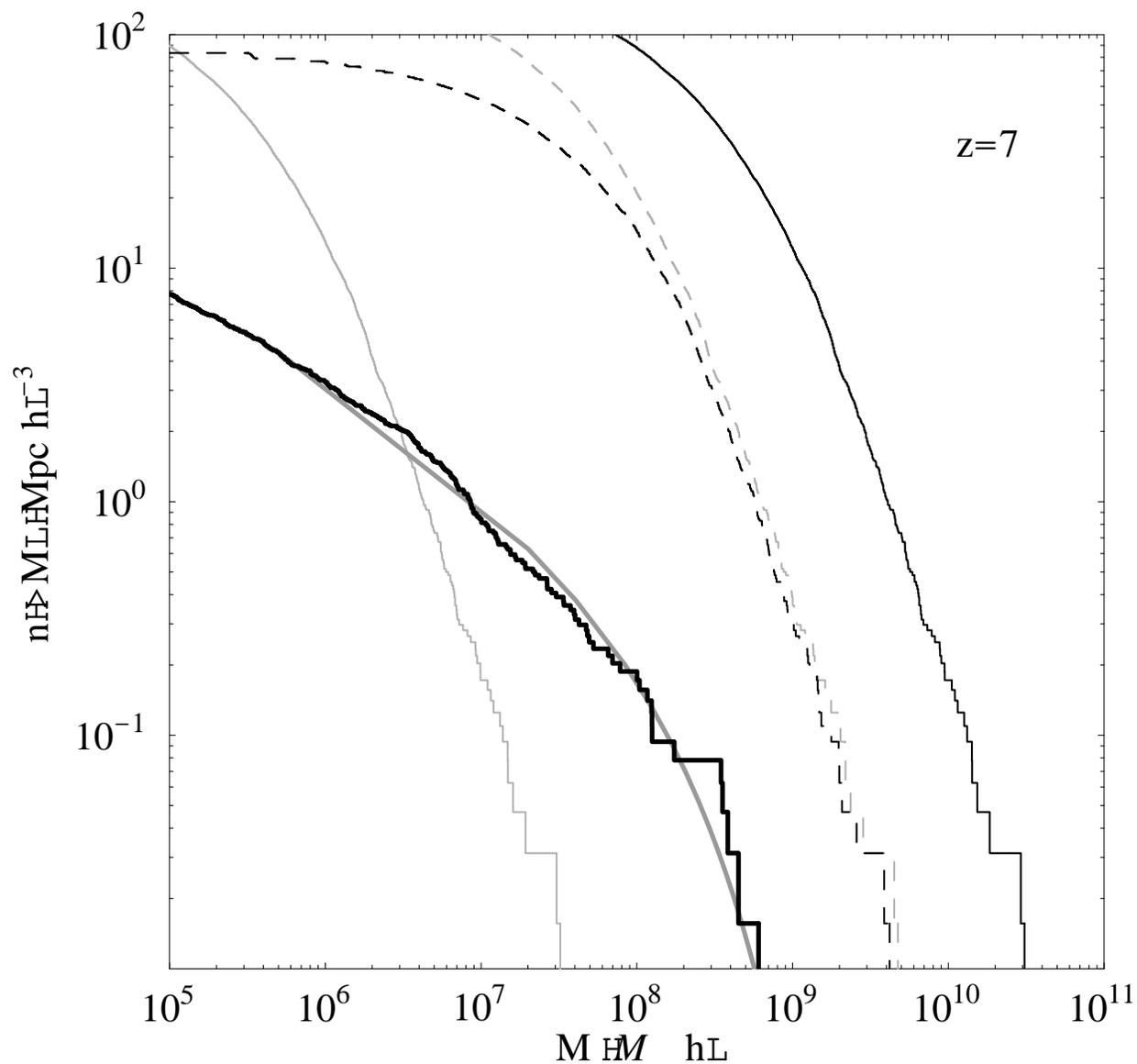}
\caption{Mass functions for bound objects 
at $z=7$, immediately after reionization.  In addition to the quantities 
in Figure \ref{fig:massfunc_one}, also 
are shown the dark matter mass function with the mass scaled by the 
ratio of stellar particles to dark matter $\Omega_{s}/\Omega_{\rm d}$ 
(thin grey), and a Schechter function fit with $\alpha \sim -1.3$ (thick grey)
to the bound stellar systems.
}
\label{fig:massfunc_two}
\end{figure}

\begin{figure}
\plotone{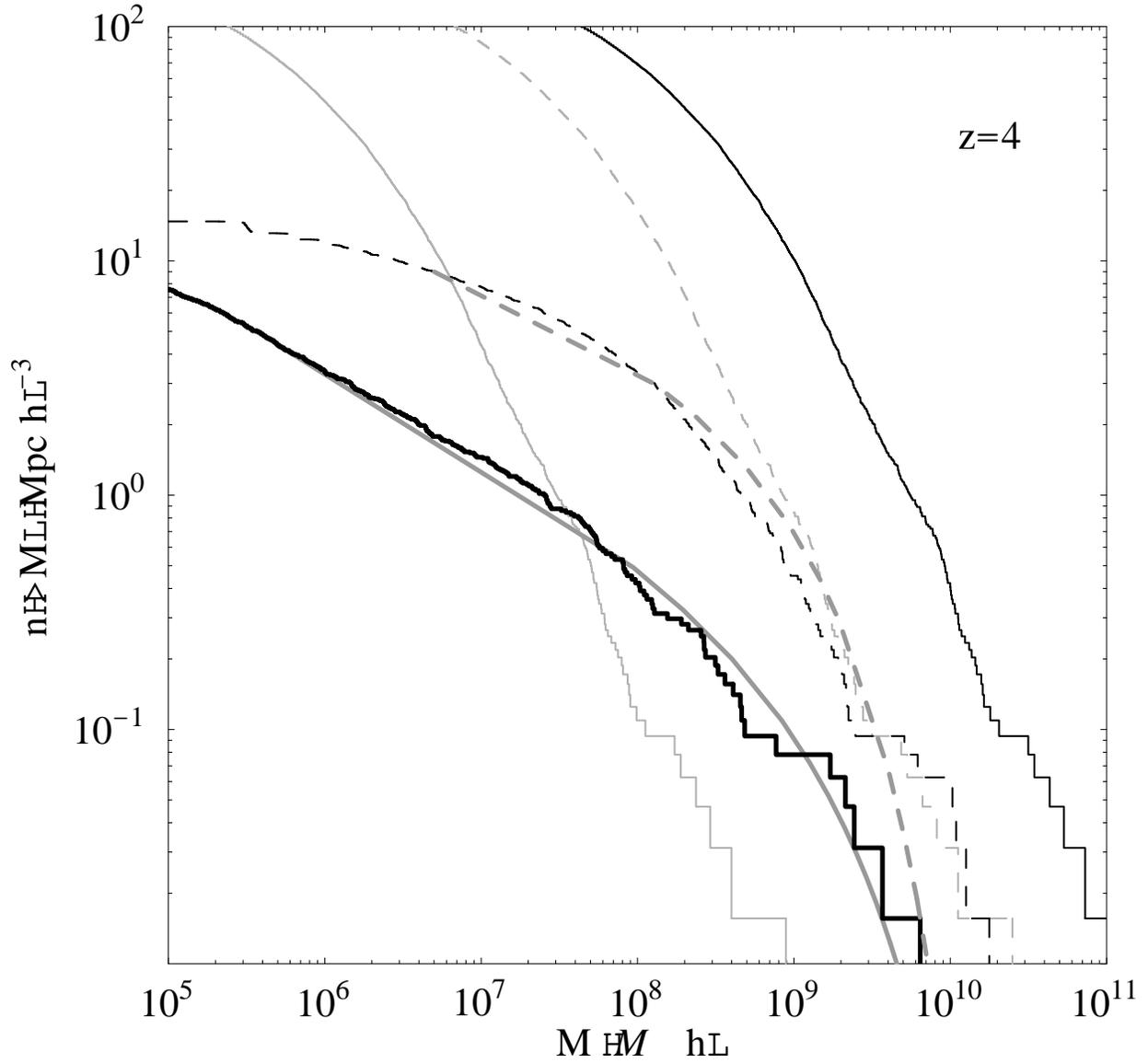}
\caption{Mass functions for bound objects 
at $z=4$.  In addition to the quantities shown 
in Figure \ref{fig:massfunc_two}, also is shown a Schechter
fit with $\alpha \sim -1.1$ (thick dashed grey) to the gas mass
function.
}
\label{fig:massfunc_three}
\end{figure}

The results of the comparison of the mass functions
are given in Figures~\ref{fig:massfunc_one}--\ref{fig:massfunc_three}.
The dark lines are measured from simulations, whereas the gray
lines are various fits.  The thin grey lines are the dark
matter mass functions scaled via equations 
(\ref{eq:numscale})--(\ref{eq:linmassscale}).
The thick gray lines are fits  with a Schechter function 
$\Phi(M)\propto\int_M^\infty m^{\alpha} \exp(-m/m_\ast) dm$
for the stellar ($z=7$ and $z=4$) and gas ($z=4$ only) components.  
The fits consistently give values of $\alpha\approx -1.2\pm 0.1$ 
when fit for objects with $M_s \geq 10^{5.7} M_\odot$ ($z\lesssim 7$) or 
$M_g \geq 10^{6.7} M_\odot$ ($z\sim 4$).  
These mass cutoffs were chosen due to numerical
resolution considerations ($1\times$ and $10\times$ 
baryonic mass resolution for stellar mass and gas mass respectively).  
The exponential cutoff $m_\ast$ may not 
be very accurate, since there is a lack of large scale power due 
to the small box size of the simulation.

\begin{figure}
\plotone{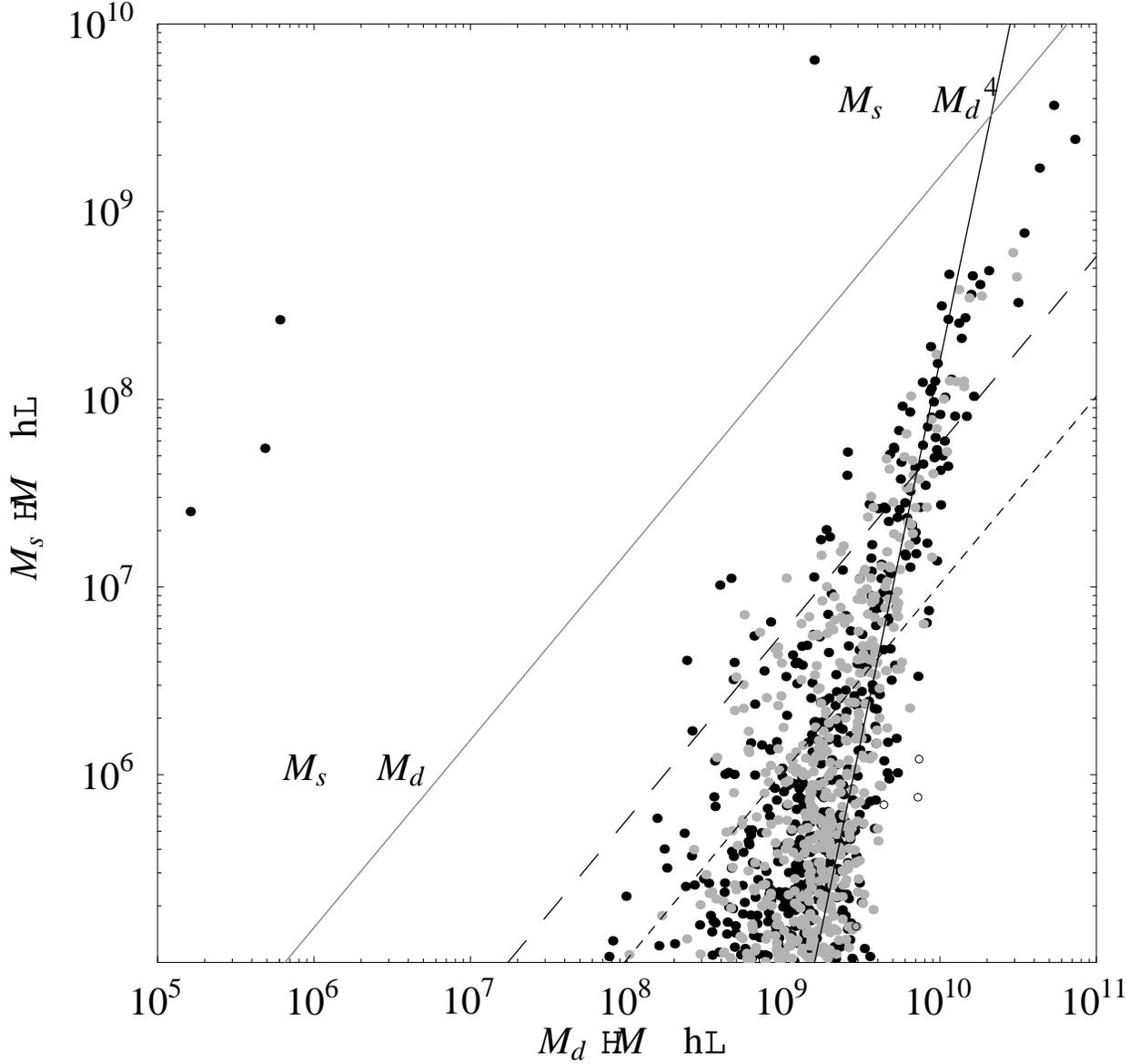}
\caption{
Scatter plot of stellar mass as a function of 
dark matter mass 
for bound objects at $z=15$ (open circles), $z=7$ (small gray points),
and $z=4$ (small black points).  The solid black line has $M_s\propto M_d^4$
corresponding to the related finding by Gnedin (2000b) that 
$M_b\propto M_{\rm tot}^4$.  The remaining three lines are linear
relations where the constants of proportionality are (a)
the universal baryon/dark matter ratio (gray) and (b) the global
stellar mass/dark matter ratios at $z=7$ (short dash), 
and $z=4$ (long dash).
}
\label{fig:mdvsms}
\end{figure}

As an additional check, we compared the bound objects' 
dark and stellar masses as shown in 
Figure~\ref{fig:mdvsms}.  Also shown are power laws
$M_s\propto M_d^4$ and $M_s\propto M_d$.  
The motivation for this $M_d^4$ power law is the finding
by \citet{G00b} that in low mass objects, heating
consequent to reionization leads
to a filtering of the baryonic mass resulting with such a relationship.
The power law $M_s\propto M_d^4$ is a generally a good fit,
while $M_s\propto M_d$ clearly fails.  

Figure~\ref{fig:mdvsms} also shows that 
the ``tail'' of the distribution around the $M_d^4$ relationship
is skewed so that for a given stellar mass $M_s$, the dispersion towards
lower $M_d$ is greater than the dispersion towards greater $M_d$.
Even given this skewed distribution, there are 
several significant outliers in the tail.  In particular, four
bound objects with stellar mass $>10^7 M_\odot/h$ 
have significantly enhanced stellar mass fractions.
For these objects, the fraction of mass in 
stellar particles is greater than even the universal
baryon/dark matter mass ratio (shown in gray).  

\begin{figure}
\plotone{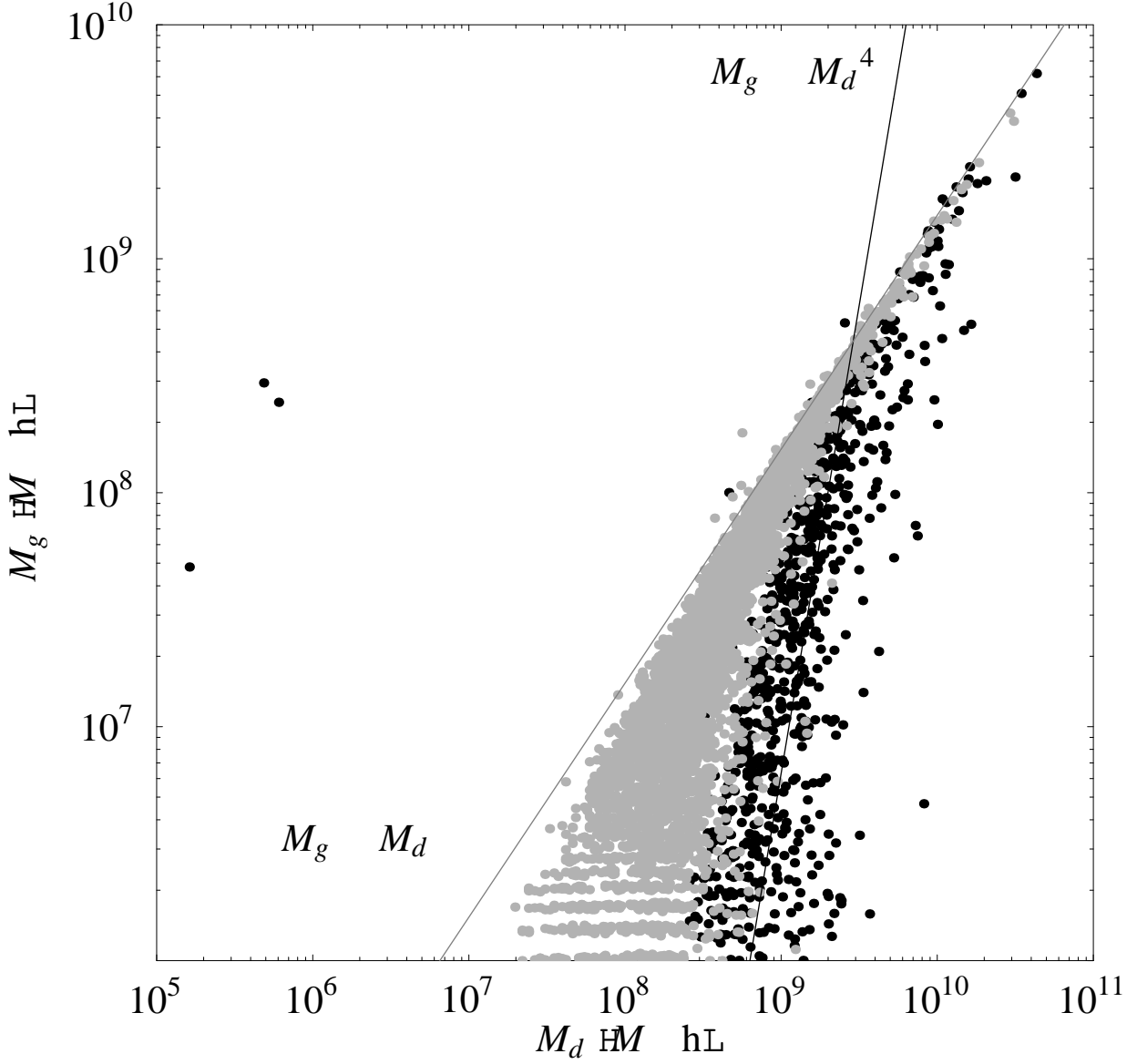}
\caption{
Scatter plot of gas mass as a function of dark matter mass 
for bound objects at  
$z=7$ (small gray points),
and $z=4$ (small black points).  The solid black 
line has $M_g\propto M_d^4$.  
The linear 
relation (gray line) has a constant of proportionality equal to
the global baryon/dark matter mass ratios.
}
\label{fig:mgvsmd}
\end{figure}

Shown in Figure~\ref{fig:mgvsmd} is the gas mass as a function
of the dark matter mass.  While the stellar mass showed a clear $\sim M_d^4$
dependence for all values of $M_s$ for all relevant epochs, 
the gas mass (consistent with 
Gnedin 2000b) shows an \emph{evolving} relationship.  The gas starts
out following the dark matter mass $M_g \propto M_d$.  After reionization,
a relation $M_g \propto M_d^4$ develops at low masses, with 
the mass at which there is a transition to $M_g \propto M_d$ 
increasing with time.

\begin{figure}
\plotone{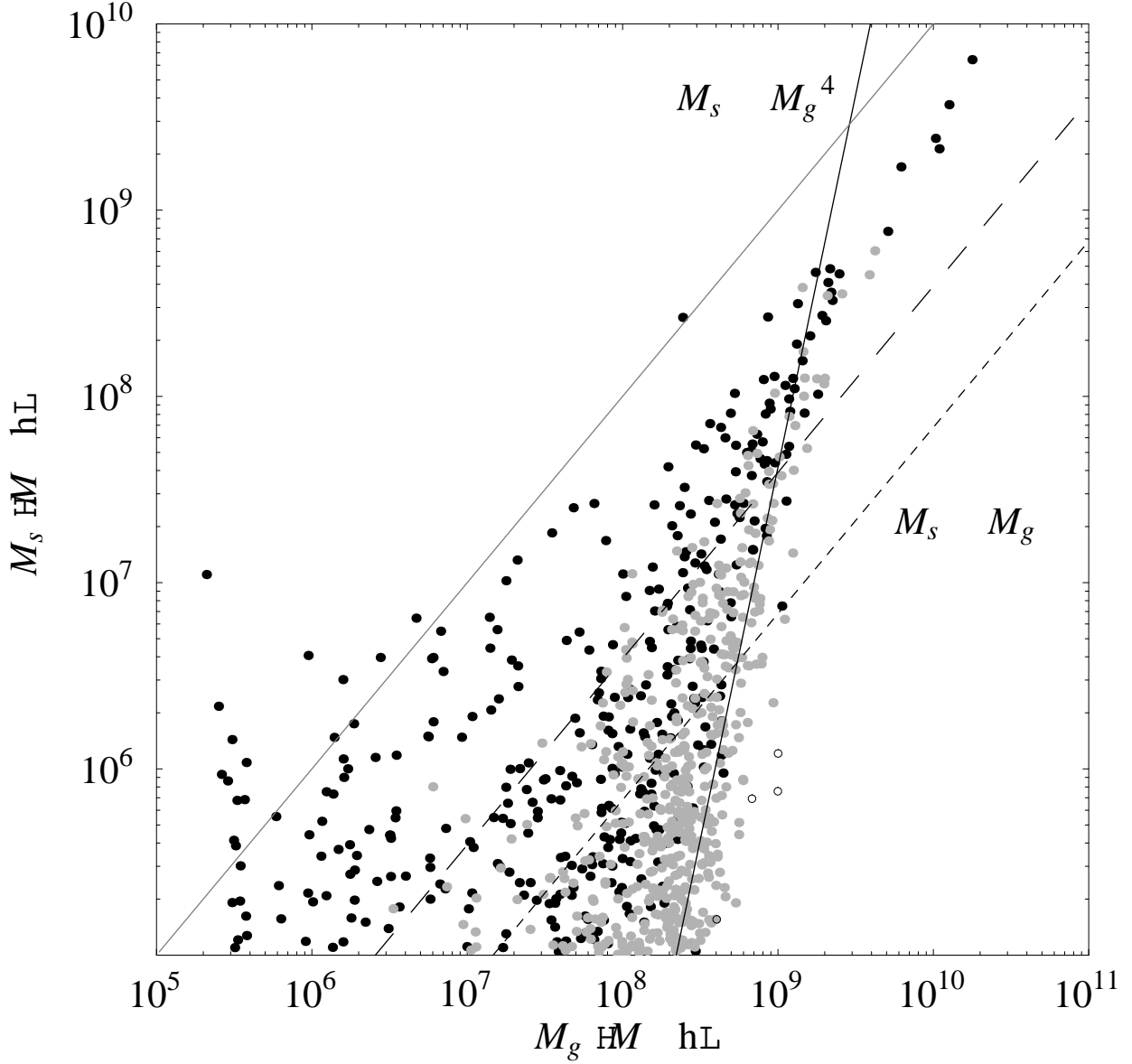}
\caption{
Scatter plot of stellar mass as a function of gas mass 
for bound objects at $z=15$ (open circles), $z=7$ (small gray points),
and $z=4$ (small black points).  The solid black 
line has $M_s\propto M_g^4$.  
The remaining three lines are linear
relations where the constants of proportionality are (a) unity (gray) 
and (b) the global stellar mass/gas mass ratios at $z=7$ (short dash), 
and $z=4$ (long dash).
}
\label{fig:mgvsms}
\end{figure}

Finally, shown in Figure~\ref{fig:mgvsms} is the 
stellar mass as a function of the gas mass of bound
objects.  Because 
the gas and dark matter are closely correlated up until $z\sim 7$,
at these epochs, the relationship between stellar mass $M_s$ and 
gas mass $M_g$ is
similar to that between the stellar mass $M_s$ and the dark mass $M_d$.  
However,
between $z=7$ and $z=4$, gas ``ejection'' from low mass 
halos is clearly evident.  This reduction in the gas mass fraction 
was noted in the analysis of \citet{G00b}. 
In some cases, the mass in gas is less
than the mass in stellar particles --- i.e., most of the gas in the 
halo has been converted to stars or ejected.

\section{Discussion: Why Is the Galaxy Mass Function Flatter
Than the Dark Matter Mass Function?}
\label{sec:discussion}

The logarithmic slope $\alpha \sim -1.2$ of the galaxy 
mass function is significantly shallower than
the dark matter halo mass function, particularly for low baryonic
masses ($M_g$ or $M_s$ $\lesssim 10^{8-9} M_\odot/h$).  Therefore 
identification of low mass halos with low mass galaxies is incorrect.  
For instance, 
a one-to-one mapping of dark matter halos to galaxies normalized
to high baryonic mass halos would result in an 
overestimate of the number of low baryonic mass systems by several orders
of magnitude. In particular, this slope 
is comparable to that observed for low mass galaxies (Cross \etal\ 2000, 
Blanton \etal\ 2000b). 

The explanation of the value of $\alpha$ is easy to explain as a 
combination of the dark matter halo mass function and the relationship
between baryonic mass and dark matter mass in halos.
The dark matter mass function follows the 
steep slope expected from the Press-Schechter \citep{PS74}
approximation
\begin{eqnarray}
n(M_d) & = &  M_d^{-2}\rho_0 \sqrt{\frac{2}{\pi}}\ 
	\frac{\delta_c}{\sigma}
	\frac{d\ln \sigma^{-1}}{d\ln M_d}
	\exp\left(-\frac{\delta_c^2}{2\sigma^2}\right) 
\nonumber\\
	& \propto & \sim M_d^{-2},
\label{eq:pressschechter}
\end{eqnarray}
where $n(M_d)$ is the comoving number density per unit mass, 
$\rho_0$ is the mean mass density today, $\delta_c$ is the critical
linear density for collapse, $\sigma$ is the rms mass fluctuation, and
we are considering small scale fluctuations where $\delta_c \ll \sigma$
and $\sigma$ is a very weak function of mass.
From Figures~\ref{fig:mdvsms} and \ref{fig:mgvsmd}, is it clear that 
\begin{equation}
M_s\propto M_d^4
\label{eq:msmdrelation}
\end{equation}
for $z \lesssim 7$ and 
\begin{equation}
M_g\propto M_d^4
\label{eq:mgmdrelation}
\end{equation}
for $z \sim 4$ at low masses.
Combining equation (\ref{eq:pressschechter}) with 
equations (\ref{eq:msmdrelation})-(\ref{eq:mgmdrelation}) 
gives
\begin{eqnarray}
n(M_s) & \propto & M_s^{-5/4} \qquad \mbox{for $z\lesssim 7$}\\
n(M_g) & \propto & M_g^{-5/4} \qquad \mbox{for $z \sim 4$}
\end{eqnarray}
for low masses.  
Thus this simple analytic derivation explains the
simulation results: 
the stellar mass function appears to be flat by $z=7$ while 
the gas mass function evolves from simple scaling
with the dark matter for redshifts of $z=15\sim 7$ to
a gradual flattening to look like the stellar mass function 
from $z \sim 7 - 4$.   

Without additional analysis, it is 
difficult to determine whether the similarity in 
the two mass functions is (a) coincidental 
--- two different mechanisms, (b) correlational --- same mechanism
of both, or (c) causal --- stars $\rightarrow$ gas.  
Clearly, though, the key to this question is understanding 
the origin of the $\sim M_d^4$ behavior for both stars \emph{and} gas, 
and how they are related.  
Whatever its origin, the result reported above appears to be robust, 
since \citet{N01} found the same
flattening in the stellar mass function in a 
simulation using a completely different 
numerical treatment of gas (e.g., Eulerian rather than Lagrangian), 
radiation, and star formation.  This work, using very 
different numerical techniques (but a comparable mass resolution) 
also found $\alpha \approx -1.2$ for the baryonic mass function.

Let us also consider the 
sensitivity to the star formation efficiency.
If the star formation prescription underestimates star formation and 
all the remaining gas should have turned into stars by $z=4$, 
the stellar mass function would retain largely the same 
shape, but with a higher normalization.   Only if the prescription
\emph{over}estimates star formation would the shape
change significantly.  In this case, the mass function would
be even flatter, and $\Lambda$CDM could have
the opposite problem of \emph{under}producing low mass 
galaxies!

In fact, assuming that \emph{all} the gas in halos at $z=4$ turns into stars 
by $z=0$ (and neglecting the effects of merging)
gives a ``turnover'' at a circular velocity 
of $\sim 40$ km/s.  That is,
the stellar mass function would follow the dark matter for $v_c \gtrsim 
40$ km/s, and would have a logarithmic slope of $\sim -1.2$ for 
$v_c \lesssim 40$ km/s.  This corresponds closely to the observed 
circular velocity distribution of galactic satellites as 
determined by \citet{Klypin00}.  Thus, if additional gas 
ejection is significant between $z=4$ and $z=0$, $\Lambda$CDM 
could indeed underproduce low mass galaxies!

\section[Conclusions]
{Conclusions}
\label{sec:conclusions}

We have analyzed the properties of the first bound luminous 
objects formed in a $\Lambda$CDM universe, as calculated
in the numerical simulation described in Gnedin (2000ab).
This simulation 
is one of only a few full-scale, non-equilibrium 
hydrodynamical simulations
to date which can \emph{individually} 
resolve the components of spheroids in the low mass range
comparable to low mass galaxies (see also Nagamine \etal\ 2001). 

We have found that the galaxy mass function at $z=4$ is
significantly flatter than the dark matter mass function, whether
one considers ``galaxies'' identified by halos' stellar particles 
or gas mass.  A flattened low mass end is found for the stellar mass function
from $z=7-4$, and for the gas mass function at $z\sim 4$.  
Taking into account the uncertainties due to the 
prescription for star formation, at $z=4$ the stellar mass function 
in a $\Lambda$CDM universe is predicted to have a low mass
logarithmic slope of $\alpha\sim -1.2$ \emph{or flatter}. 
The reason for these shallower baryonic mass functions 
is a combination of two effects: 
(a) the dark matter halo (differential) mass function 
is consistent with the Press-Schechter
formulation at low masses of $n(M) \propto M^{-2}$ and
(b) heating/cooling and ionization processes appear to cause 
baryons to collect in halos with the relationship $M_b \propto M_d^4$
at low masses.  Combining (a) and (b) gives $n(M_b) \propto M_b^{-5/4}$,
comparable to the simulation results. 
Thus, the ``underabundance'' of low luminosity galaxies 
relative to halos emerges naturally from these results,
implying that no ``new physics'' is necessarily required to understand
the low mass end of the galactic mass function.  
However, because gravitational clustering is still evolving between 
$z=4$, when the simulation ends, and the present epoch, 
it remains to be seen whether $\Lambda$CDM with the appropriate 
astrophysics can be reconciled with the 
observed clustering properties of low mass galaxies.

\acknowledgments

Discussions with Ken Nagamine are gratefully acknowledged.
This work was supported in part by NSF Grant AST-9318185.

\clearpage


\clearpage 

\clearpage

\begin{deluxetable}{ll}
\tablecaption{Fraction of Mass in Stellar Particles
\label{tab:meanmassfrac}}
\tablewidth{0pt}
\tablehead{
\colhead{Redshift} & \colhead{$f_{s}$}}
\startdata
15 & $6\times 10^{-7}$  \\
7 & $9\times 10^{-4}$  \\
6 & $0.002$  \\
5 & $0.003$  \\
4 & $0.005$  \\
\enddata



\end{deluxetable}

\end{document}